\documentclass[showpacs,prl,amsmath,amssymb,aps,floatfix]{revtex4}
\usepackage{graphicx}
\usepackage{dcolumn}
\usepackage[sort&compress]{natbib}
\usepackage{bm}
\usepackage{color}
\usepackage[dvips]{epsfig}
\newcommand{\comment}[1]{}
\newcommand{\AAA}{{\rm \,\AA}}

\begin{document}

\title{Thermal-induced proteinquake in oxyhemoglobin}


\author{S. G. Gevorkian$^{1,3}$, A.E. Allahverdyan$^{2,3}$, D.S. Gevorgyan$^{4}$,
Chin-Kun Hu$^{1}$\footnote{E-mail: huck@phys.sinica.edu.tw.}}

\affiliation{$^{1}$Institute of Physics, Academia Sinica, Nankang,
Taipei 11529, Taiwan}

\affiliation{$^{2}$ Laboratoire de Physique
Statistique et Syst\`emes Complexes, ISMANS, 44 ave. Bartholdi,
72000 Le Mans, France}

\affiliation{$^{3}$Yerevan Physics
Institute, Alikhanian Brothers St. 2, Yerevan 375036, Armenia}

\affiliation{$^{4}$ Institute of Fine Organic Chemistry, 26
Azatutian ave., Yerevan 0014, Armenia}

\date{\today}

\begin{abstract}
  Oxygen is released to living tissues via conformational changes of
  hemoglobin from R-state (oxyhemoglobin) to T-state
  (desoxyhemoglobin). The detailed mechanism of this process is not
  yet fully understood. We have carried out micromechanical
  experiments on oxyhemoglobin crystals to determine the behavior of
  the Young's modulus and the internal friction for temperatures
  between 20 C and 70 C. We have found that around 49 C oxyhemoglobin
  crystal samples undergo a sudden and strong increase of their
  Young's modulus, accompanied by a sudden decrease of the internal
  friction. This sudden mechanical change (proteinquake) takes place
  in a partially unfolded state and precedes the full denaturation
  transition at higher temperatures. The hemoglobin crystals after the
  proteinquake has the same mechanical properies as the initial state
  at room temperatures. We conjecture that it can be relevant for
  explaining the oxygen-releasing function of native oxyhemoglobin
  when the temperature is increased, e.g. due to active sport. The
  effect is specific for the quaternary structure of hemoglobin, and
  is absent for myoglobin with only one peptide sequence.

\end{abstract}
\maketitle

\begin{figure}
\centerline{\includegraphics[width=.4\textwidth]{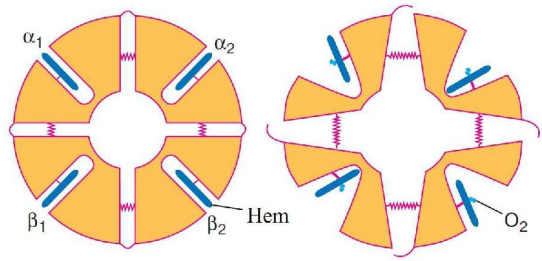}}
\vskip 2 mm \caption{(Color online) Schematic representation of the
hemoglobin quaternary structure with four globular sub-units (yellow
color: $\alpha_1$, $\alpha_2$, $\beta_1$, $\beta_2$) and four heme
groups (blue color). (Left) T-form, where the access to the oxygen
binding parts of the heme groups is restricted and the overall
structure is more compact; (Right) R-form with easier access to the
heme groups by oxygen molecule ${\rm O_2}$ and a more loose
structure. The cooperativity of hemoglobin during oxygen binding is
related to the transition from the T form to the R
form.} \label{fig00}
\end{figure}

Since its discovery in 1840, the hemoglobin is one of the most
extensively studied proteins \cite{perutz,eaton}. This is related to
its important physiological function: it carries oxygen from the lungs
throughout the body allowing us to breathe and live. It consists of
four globular units linked into a double-dimer tetrameric structure
\cite{perutz,eaton} as shown schematically in Fig.~1. Each unit can
carry one oxygen molecule ${\rm O_2}$ attached to its heme group. The
hemoglobin structure is adapted to the needs of its function. First,
its oxygen binding is cooperative: the response of hemoglobin with
respect to ${\rm O_2}$ concentration has a S-shaped region, thereby a
relatively slight decrease of the oxygen concentration between the
lungs and the body brings in a significant decrease of the bound
oxygen (up to 25\%) \cite{perutz,eaton}. Second, the oxygen binding
ability decreases upon reducing the pH factor or increasing the
concentration of ${\rm CO}_2$ \cite{bohr}. Due to this Bohr's effect
\cite{bohr} a tissue with a stronger need of oxygen receives it
more. Cooperative oxygen unbinding of hemoglobin is explained by the
change of its tetrameric conformational structure induced by binding
of the first oxygen molecule \cite{eaton}; see Fig.~\ref{fig00}.
The conformational states with (R-state or oxyhemoglobin) and without
(T-state or desoxyhemoglobin) the four oxygen molecules are clearly
distinguishable by their shape, as seen via X-ray crystallography
\cite{perutz,eaton}. {\it In vivo} those states are more like dynamic
ensembles than fixed conformations \cite{lukin_nmr_structure}.

Many aspects of hemoglobin are well understood by now
\cite{eaton}. However, the physics of conformational changes and
their interaction with external factors (pressure, oxygen
concentration, temperature) is still under active scrutiny
\cite{swed,yan,artmann,ArtmannGM}. In particular, this concerns the thermal
response of hemoglobin that is traditionally studied via
denaturation experiments. It is known that the multidomain
hemoglobin does not unfold via a single transition \cite{swed}.
Rather, there is a wide transition zone $\simeq 40^\circ -
60^\circ$ C that includes several events \cite{swed}. These events
are typically interpreted via the concepts of unfolding and/or
internal aggregation, two standard mechanisms normally applied for
describing thermal responses of multi-subunit proteins
\cite{swed,yan}.

Here we report results of micromechanical experiments carried out
on crystals of horse and human hemoglobin. We show that in its
partially unfolded state|i.e. for a temperature higher than the
physiological temperatures, but lower than the unfolding
temperature|the hemoglobin responds to heating by a sudden release
of force and a subsequent jump of the Young's modulus, which is similar to
proteinquake reported by Ansari et al. \cite{frau}.
The detailed structure of this effect is different
for human and horse hemoglobin. We argue that the effect relates
to certain slowly relaxing degrees of freedom of the quaternary
structure that accumulate energy during heating and then suddenly
release it at a well-defined temperature that is specific for
hemoglobin.

Such an effect is absent in the thermal response of myoglobin.
This is a single-unit globular protein that displays a visible
transition towards a partially denaturated state around a certain
unfolding temperature \cite{swed}. Myoglobin also binds and
unbinds oxygen, but does so without a sizable cooperativity. This
relates to its function: myoglobin is a depot (not transporter) of
oxygen in muscles. The difference between hemoglobin
and myoglobin was also observed in
heating and re-cooling data of Young's modulus for
two systems to be reported below.



The  proteinquake  reported here for hemoglobin
crystals has several predecessors in biopolymer physics. Ansari et
al. found an indirect experimental evidence that the
low-temperature ligand unbinding of myoglobin is modulated by a
sudden release of energy accumulated due to the ligand binding
\cite{frau}. They proposed the term {\it proteinquake} for such
effects. Later on, it was found computationally that proteinquakes
are relevant for the functioning of myosin \cite{tol} and
adenylate kinase \cite{wol}. More specifically for hemoglobin, it
is known that when crystals of deoxyhemoglobin are exposed to
oxygen, they shatter due to the force released during the
conformational transition from deoxy to oxyhemoglobin
\cite{53,bourg}. Postponing detailed connections with literature
till the end of this paper, we stress already here that the
presented report seems to be the first one, where a proteinquake
effect was found under heating which is generally supposed to
diminish mechanical features of biopolymers.

\begin{figure}
\centerline{\includegraphics[width=.4\textwidth]{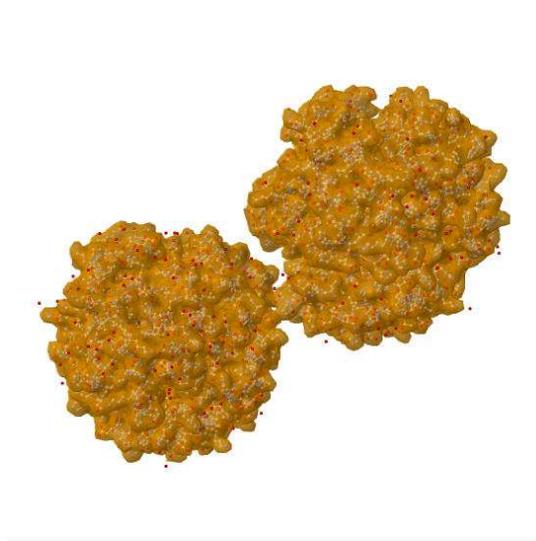}}
\caption{(Color online) Contacts between two deoxyhemoglobin
molecules in monoclinic crystal at resolution $1.74\AAA$ with
parameters ${\rm P}(1~~2_1~~1)$, $a=63.15 \AAA$, $b=83.59 \AAA$,
$c=53.80 \AAA$, $\alpha=90.0^\circ$, $\beta=99.3^\circ$,
 $\gamma=90.0^\circ$ \cite{perutz3,database}. Red points represent
 water molecules from the first Langmuir adsorption layer. The
 relative humidity is 95\%-98\%. } \label{fig0}
\end{figure}

Note in this context that the advantage of using biopolymer
crystals for measurement is that there is a possibility of
controlling and displaying|via intermolecular contacts regulated
by the crystal syngony and the water content|those motions of the
macromolecule that can have only transient character in the
solution \cite{Starikov,morozov_gevorkian}. Almost all the basic
information on the hemoglobin structure came from experiments on
soild-state hemoglobins. Figure \ref{fig0} illustrates the
structure of the hemoglobin crystal \cite{perutz,eaton,perutz3,database}.
The solid state hemoglobin is closer to its {\it in vivo} state in
mammal erythrocytes, where the hemoglobin is densely packed with
concentration $\simeq 34\%$ \cite{trincher}.




\vskip 3 mm

\noindent{\large \bf Results}

\begin{figure}
\centerline{\includegraphics[width=.4\textwidth]{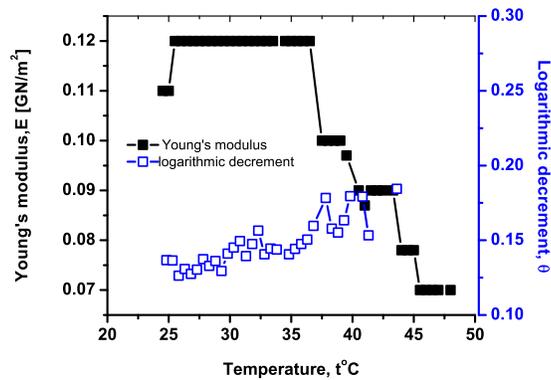}}
\caption{(Color online) The Young's modulus and the logarithmic
decrement of damping of the human hemoglobin under heating. The
relative humidity is now $95\%$.} \label{fig3}
\end{figure}

\noindent{\bf 1.} We start with the denaturation curve of the human
oxyhemoglobin at relative humidity $95\%$ as shown in Fig.~\ref{fig3}. The
Young's modulus $E$ is stable between $25^\circ$C and $36^\circ$C
and then decreases in discrete steps. In between of each step $E$
is constant. We argue below that that these discrete steps are
related to partial dissociation of the quaternary structure. The
measured internal friction $\theta$ increases, because the
structure breaking liberates new degrees of freedom that are able
to dissipate the energy of forced oscillations; see
Fig.~\ref{fig3}.

But we could not perform the experiment for temperatures higher
than $49^\circ$C, because (at the employed excitation frequency
$\simeq 5$ kHz of the plate oscillations) the crystals clove to
pieces. The measurement of $\theta$ had to be terminated even
earlier due to instability of results; see Fig.~\ref{fig3}. The
same breaking effect at $49^\circ$ C was observed for crystals of
horse hemoglobin prepared under the same relative humidity $95\%$.
This effect indicates on conformational changes taking place in
the hemoglobin macromolecule. Recall in this context that
deoxyhemoglobin crystals shatter after exposure to oxygen, since
they cannot support a large conformational change related to the
transition from deoxy to oxyhemoglobin \cite{53,bourg}. The
difference with our situation is that we work with oxyhemoglobin
and that with us the crystals shatter after heating.

\begin{figure}
\centerline{\includegraphics[width=.4\textwidth]{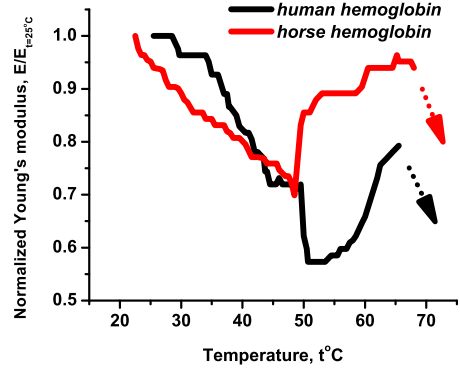}}
\caption{(Color online) The Young's modulus of the horse versus
human hemoglobin. Both samples were prepared at the same initial
temperature $t=25^\circ$ and relative humidity $75\%$. The Young's
modulus at temperature $t=25^\circ$: $E_{t=25^\circ}=0.75\,{\rm
GN/m}^2$ for the horse hemoglobin sample and
$E_{t=25^\circ}=0.94\,{\rm GN/m}^2$ for the human sample. }
\label{fig4}
\end{figure}

\noindent{\bf 2.} To understand this effect, we equilibrated our samples
under a smaller relative humidity of $75\%$ at $25^\circ$ C. This
was expected to prevent strong instabilities in the crystal
\cite{biswal}, since inter-molecular contacts in the crystal get
stronger due to less inter-molecular water (Fig.~\ref{fig0}).
Figure \ref{fig4} displays the behavior of the Young's modulus
for the human and horse hemoglobin under heating at relative
humidity $75\%$. Note that compared to the $95\%$-humidity, the
absolute value of the Young's modulus increased for about $8$
times; cf Fig.~\ref{fig4} with Fig.~\ref{fig3}.

For temperatures from $20^\circ$ C till $49^\circ$ C the Young's
modulus $E$ of the human hemoglobin decreases again in discrete
steps. However, due to less inter-molecular water these steps are
now shorter. Indeed, consider the temperature interval $25^\circ
-36^\circ$ C. For $95\%$-humidity $E$ is constant there [see
Fig.~\ref{fig3}], but for $75\%$-humidity it still makes one
sudden change in this interval; see Fig.~\ref{fig4}. Hence the
inter-molecular water does stabilize the hemoglobin structure
against dissociation, though it decreases the absolute value of
$E$. For horse hemoglobin the decay of $E$ for temperatures from
$20^\circ$ C till $49^\circ$C is more gradual, but the stepwise
change is still visible; see Fig.~\ref{fig4}.

\begin{figure}
\centerline{\includegraphics[width=.4\textwidth]{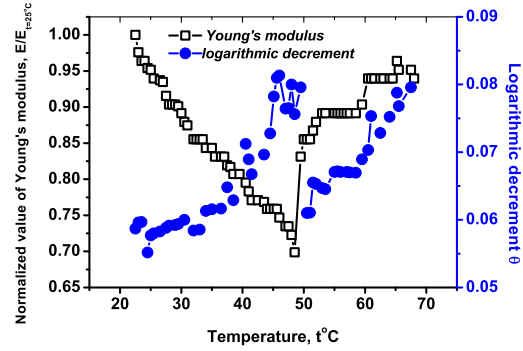}}
\caption{(Color online) The normalized Young's modulus
$E/E_{t=25^\circ}$ and the logarithmic decrement of damping,
$\theta$, versus temperature $t$ for monoclinic crystals of horse
hemoglobin. The relative humidity is $75\%$. The heating rate is
$0.1^{\circ}\,{\rm C/min}$ and $E_{t=25^\circ}=0.75\,{\rm
GN/m}^2$. } \label{fig5}
\end{figure}

\begin{figure}
\centerline{\includegraphics[width=.4\textwidth]{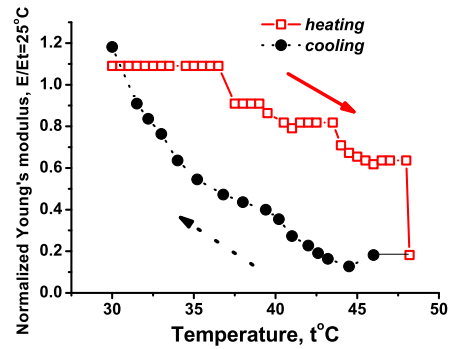}}
\caption{(Color online) The Young's modulus of the human
hemoglobin under heating and re-cooling in the vicinity of
$49^\circ$ C. The relative humidity is $95\%$. The heating rate is
$0.1^{\circ}\,{\rm C/min}$. Temperature is smaller than $49^\circ$
C and $E_{t=25^\circ}=0.11\,{\rm GN/m}^2$. } \label{fig6}
\end{figure}

\begin{figure}
\centerline{\includegraphics[width=.4\textwidth]{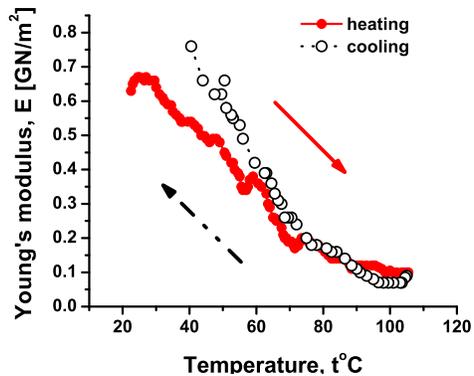}}
\caption{(Color online) The Young's modulus $E$ versus temperature
$t$ under heating and recooling for
           monoclinic crystals of sperm-whale myoglobin. The denaturation
           process is seen to be approximately reversible.
} \label{fig7}
\end{figure}

\begin{figure}
\centerline{\includegraphics[width=.4\textwidth]{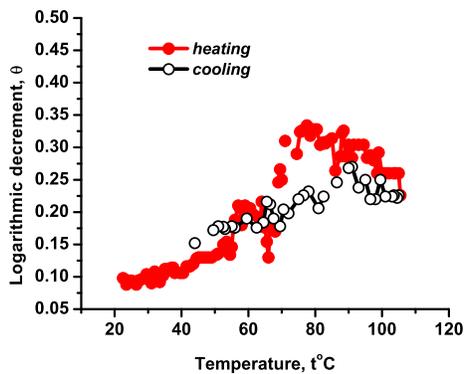}}
\caption{(Color online) The logarithmic decrement of damping,
$\theta$ versus temperature $t$ under heating and recooling for
monoclinic crystals of sperm-whale myoglobin. } \label{fig8}
\end{figure}

\noindent{\bf 3.} But the temperature $49^\circ$ C is again a special one
both for the human and horse hemoglobin; see Figs.~\ref{fig4},
\ref{fig5} and \ref{fig6}. In its vicinity, the Young's modulus $E$ changes {\it
abruptly}. We prescribe these effects to the quaternary structure
of hemoglobin, since myoglobin which lacks this structure, but
still has well-defined tertiary and secondary structure does not
show this effect [see Fig. \ref{fig7} and Fig. \ref{fig8} below]. Thus, it is
plausible that certain degrees of freedom of the quaternary
structure have long relaxation times. During gradual heating, they
go out of equilibrium, accumulate energy in elastic deformations
and then suddenly release this strain energy at $49^\circ$ C
(proteinquake).

The abrupt change of the Young's modulus for relative humidity $75\%$
is clearly the same effect that is responsible for breaking the
crystals at a larger relative humidity $95\%$, where
inter-molecular contacts are weaker [see {\bf 1}]. However, for
the human hemoglobin $E$ abruptly {\it decreases} at $49^\circ$ C,
while for the horse hemoglobin it abruptly {\it increases}.  We
see that the precise type of the mechanical event|i.e., whether
its Young's modulus increases or decreases|depends on the type of
hemoglobin (human or horse), but the temperature of the event is
to a larger extent independent of the type, as Fig.~\ref{fig4}
shows [we come back to this difference in {\bf 4}]. It is also
independent from the relative humidity, in contrast to the
absolute value of $E$ and the pattern of its change.

Figure~\ref{fig5} presents the behavior of both $\theta$ and $E$
for the horse hemoglobin. It is seen that in the immediate
vicinity of $49^\circ$ C, $\theta$ abruptly decreases basically to
the value it had at $\simeq 35^\circ$ C. This indicates that the
degrees of freedom liberated during the previous melting stage get
blocked again.

The evolution of $E$ and $\theta$ under heating with temperatures
higher  than $50^\circ$ C is different: $\theta$ starts to grow
again indicating a new trend in structure breaking; see
Fig.~\ref{fig5}. But $E$ gradually increases for both human and
horse hemoglobin; see Fig.~\ref{fig4}.  We interpret this effect
via the intra-tetrameric aggregation of partially denaturated
monomers of hemoglobin, because such increase of $E$ is absent for
the monomeric myoglobin; see Fig.~\ref{fig7} below.

It is impossible to prescribe the event at $49^\circ$ C to the
aggregation {\it only}, because the changes of $E$ and $\theta$
take place within a too narrow temperature interval.  We also
cannot prescribe this effect to unfolding, since the decreasing
$\theta$ indicates on ordering (rather than disordering) of
certain degrees of freedom. Apparently, the only possibility left
is that the event at $49^\circ$ C indicates on the transition of
the partially denaturated tetramer from the R-state to another
conformational state with different visco-elastic features.

Figure~\ref{fig6} displays the heating-recooling dynamics of
the human hemoglobin at  relative humidity $95\%$ around $49^\circ$ C. The effect is
irreversible, but the initial value of $E$ is roughly recovered
for $30^\circ$ C, albeit the characteristic stepwise pattern of
decreasing $E$ under heating is not seen during the recooling.
This indicates that the discrete steps relate to dissociation of
the hemoglobin quaternary structure, which (as compared to the
tertiary and secondary structures) is expected to be the most
fragile one. Figure~\ref{fig6} confirms that the event at
$49^\circ$ C takes place in a partially unfolded state.

\noindent{\bf 4.} We now face a non-trivial situation: the event happens at
the same temperature $49^\circ$ C for both horse and human
hemoglobin. But, the behavior of the Young's modulus is very
different: it suddenly increases for the horse hemoglobin (more
rigid structure for higher temperatures), but decreases for human
hemoglobin (less rigid). We repeated the experiment with several
different samples, to be sure that the difference between human
and horse hemoglobin is well reproduced.

It is well known that these two macromolecules differ by 20-25
acid residues on each hemoglobin unit \cite{clegg}. Due to such
differences the horse hemoglobin has lower reactivity with respect
to certain cofactors regulating oxygen binding \cite{giardina}.

We conjecture that these biochemical differences are reflected in the
mechanical features of hemoglobin around $49^\circ$ C. The slow
degrees that are responsible for the mechanical event at $49^\circ$ C
are most probably the same for the horse and human hemoglobin,
otherwise the temperature $49^\circ$ C could not be the same for both
situations. But since the environments of these degrees of freedom in
horse and human hemoglobin are different, we get a different overall
response for horse and human situations.



\noindent{\bf 5.} For temperatures higher  than $65^\circ$ C the Young's
modulus for the hemoglobin crystals [both for human and horse]
abruptly decreases again (not shown on figures) indicating on its
full denaturation. For such high temperatures we expect that even
the secondary structure of the macromolecule is broken.  Our
experimental samples became unstable for temperatures higher than
$70^\circ$ C, so that no reliable data could be extracted.



\noindent{\bf 6.} To gain more evidence on whether the described effect is
specific for the tetrameric structure of hemoglobin, we performed
the same experiment with the myoglobin crystals and showed the result in
Fig.~\ref{fig7}. Upon heating, the Young's modulus $E$ of the
myoglobin decreases {\it gradually}, without discrete steps.  This
is consistent with breaking the tertiary structure of the
myoglobin towards acquiring a more labile state. This also
confirms that the discrete steps seen in Figs. \ref{fig3} and
\ref{fig4} relate to the dissociation of the quaternary
structure, which is absent for myoglobin. Apart from relatively
small non-monotonicity of $E$ around $57^\circ$ C, the behavior of
$E$ is monotonic and reversible upon recooling. It is seen though
that the reversibility is better visible for $T>57^\circ$ C than
for lower temperatures, since for low temperatures $E$ is more
influenced by the well-defined myoglobin tertiary structure, which
is not completely recovered after recooling.  It is known that the
calorimetric experiments for myoglobin show a well-displayed
transition to a partially denaturated (molten) state, where its
tertiary structure is partially lost; see \cite{swed} for further
references.  Depending on certain experimental conditions this
happens around $60^\circ-80^\circ$ C \cite{swed}. But the
mechanical features of myoglobin do not change suddenly during
this transition to the partially molten state. This corresponds to
the behavior of the logarithmic decrement of damping for the
myoglobin|see Fig.~\ref{fig8}-- and is consistent with the
gradual decrease of $E$ with temperature; see Fig.~\ref{fig7}.


{\it Relations to previous work.} Several experimental studies carried
out via various methods (calorimetry, optics) on liquid-state samples
displayed that something peculiar happens with hemoglobin around
$49^\circ$ C. \cite{swed,yan,artmann}. The authors of \cite{swed} and
\cite{yan} prescribed the $49^\circ$ C event to the onset of
aggregation. As we saw, this is correct, but essentially incomplete:
the aggregation indeed starts around $49^\circ$ C, but there is
certainly more there than simply aggregation. Artmann {\it et al.}
found indications of conformational transitions (not reducible to
aggregation) at $49^\circ$ C using optical methods on liquid state
human hemoglobin \cite{artmann}. This is close to our results. They
extensively studied hemoglobin of other species and noted that the
event correlates with the physiological temperature and relates to the
motion of erythrocytes (red blood cells) \cite{artmann}. They also suggested that the
scenario of this conformational transition can be similar to the glass
transition in polymers \cite{artmann}. Also this suggestion is
confirmed by our results, because the peak of the internal friction
around transition temperature [see Fig.~\ref{fig5}] is a known
indication of the glass transition, as was employed recently for
detecting the glass transition in collagen \cite{collagen,13PlosOne};
see also \cite{morozov_gevorkian,iben,10MaHua,10MaHub,10HuMa} in the
context of glassy features of biopolymers.

Several decades ago it was conjectured that conformational changes
related to biopolymer functioning proceed via mechanical motion of
certain mesoscopic degrees of freedom \cite{cherna}. This {\it
protein-machine} conjecture was taken up in \cite{frau}, where
certain aspects of the low-temperature kinetics of myoglobin were
interpreted via proteinquakes: a sudden release of energy
accumulated in elastic degrees of freedom. Later on, proteinquakes
were found to be relevant for functioning of myosin \cite{tol} and
adenylate kinase \cite{wol}. An important aspect of this research
was that proteinquakes were connected to partial unfolding of the
biopolymer tertiary structure \cite{tol,wol}. This agrees with our
finding that the heating-induced proteinquake at $49^\circ$ C
takes place in a partially unfolded state of hemoglobin.

Meanwhile the protein-machine conjecture was supported from
another angle: not only the protein functioning resembles that of a
machine, but also the performance of an optimal (high efficiency and a
large power) heat engine|as described by a generalized Carnot cycle|has
shown deep analogies with protein physics and the
folding-unfolding transition \cite{carnot}.


\vskip 5 mm

\noindent{\large \bf Discussion}

Our main message is that upon temperature elevation
horse and human hemoglobins experience a conformational transition
around $49^\circ$ C detectable via suddenly changing Young's
modulus and decreasing internal friction.  We argued that this is
a mechanical event and that it cannot be traced back to
denaturation and/or aggregation. The precise scenario of the
event|but not its temperature|appears to be dependent on the type
of hemoglobin. We conjectured that this difference relates to
structural differences in the horse and human hemoglobin.


The message of our results for the hemoglobin
functioning is that its mechanical features can be triggered by
temperature {\it in addition} to other pertinent factors such as
pH or oxygen concentration. In particular, our results can turn
out to be relevant for understanding the process of oxygen release
that in tissues which strongly need oxygen should take place much
faster than the oxygen consumption in lungs. During the {\it in
vitro} experiments we naturally do not see those additional
factors and thus the temperature has to be larger than the
physiological range for seeing the effect.

Note that the global temperatures $\geq 42^\circ$ C are lethal for
humans and horses. However, the {\it local} temperature can easily go
to values larger than $42^\circ$ C without causing any serious trouble
for living tissues. A pertinent example is the hyperthermia treatment
of cancer, where the local temperature goes to values larger than
$45-48^\circ$ C without destroying healthy tissues. Note as well that
the orientation of hemoglobin molecules in erythrocytes is not random
(in that respect they are similar to protein crystals). First
indications of this fact were obtained some 30 years ago, but were not
widely disseminated in that time \cite{fok}. More recently this fact
was confirmed by several independent studies; see e.g. \cite{fokfok}.


In conclusion, we reiterate that the main experimental result of this
work|proteinquake in heated hemoglobin|can be interpreted as a sudden,
temperature-controlled mechanic motion. The mechanic character of this
motion is to be compared with the fact that certain features of
proteins may well have glassy features. Recall that the glassy state
is yet another (different from a mechanic motion) form of
non-equilibrium that is characterized by slow relaxation and strong
memory effects \cite{79prbSG,13aipHu,14cjpHu}. By now there are many
experimental and numerical indications of such a state in biopolymers
\cite{morozov_gevorkian,collagen,13PlosOne,iben,10MaHua,10MaHub,10HuMa}. It
should be interesting to carry out direct experiments and study in
which specific way the glassy state in proteins coexists with the
mechanic motion. 




\vskip 5 mm

\noindent{\large \bf Methods.}

\noindent{\bf Materials.}
Monoclinic crystals of horse and human hemoglobin
were grown following the modified method of Drabkin
\cite{drabkin}.
The solution contained $33$ mg/ml oxyhemoglobin,
$1\%$ sodium oxalate, $0.9$ M potassium phosphate and had ${\rm
pH}=6.5$. It was poured on glass weighing cups, which were put
inside the crystallization chamber. At the bottom of the chamber
we poured a solution of $3\%$ sodium oxalate and $2.7$ M potassium
phosphate (${\rm pH}=6.5$). Crystallization was processed under
$4-6^\circ$ C for several months. The crystals came out as thin
rhomboids with the length around $L=4$ mm, width $b=0.05$ mm and
thickness $h=0.005$ mm (we stress that different crystal samples
have slightly different values of $L$, $b$ and $h$, what are given
here are the characteristic values). This form did not differ from
that described in \cite{drabkin} where the crystallization was
processed in the presence of sodium oxalate and $\left ({\rm NH}_4
\right)_2{\rm SO}_4$. The obtained crystals can be utilized
directly (without fixation) for our micromechanical experiments.
A typical structure is shown in Fig. 2 \cite{perutz3,database}.
Monoclinic crystals of sperm-whale myoglobin were grown following
the method of \cite{kendrew}.



\noindent{Experimental methods.} The  dynamic Young's modulus $E$ and the
logarithmic decrement of damping $\theta$ for hemoglobin and
myoglobin crystals were studied via the method described in
\cite{Mor2,morozov_gevorkian}. The method is based on analyzing
electrically induced  transverse mechanical oscillations of the
plate which is fixed by a cantilever.  The dynamic Young's modulus
measures the elasticity degree. To measure $E$, one changes
smoothly the frequency $f$ of the induced oscillations to
determine experimentally the basic resonance frequency $f_0$,
which corresponds to the maximal oscillation amplitude of plate's
loose  end. The Young's modulus for the long axial direction of
the plate is given as \cite{Lan}
\begin{equation}
E=3.19\,\frac{f_0^2L^4\rho S}{I},
\end{equation}
where $L$ is the length of sample, $S=bh$ is the cross-section
area of the plate, $\rho$ is its density, and $I={b h^3}/{12}$ is
the main inertia moment of the plate.


The logarithmic decrement of damping $\theta$ is used as a measure
of the internal friction, and is defined as
$\theta=\ln[{A(t)}/{A(t+T)}]$, where $A(t)$ is the oscillation
amplitude in time $t$, and $T$ is the period.
 $\theta$ is measurable by two related methods. One can measure
the length of the resonance curve and calculate $\theta$ according
to: $\theta={\pi\Delta f}/{f_0}$, where $\Delta f$ is the
difference of frequencies between the oscillation amplitude at the
maximum amplitude and $\sqrt{2}$ times less than the maximum. A
more precise determination method amounts to measuring the phase
shifts between the oscillations of exciting force and the sample
loose end \cite{Mor2,morozov_gevorkian}.





\vskip 6 mm

\noindent{\large \bf Acknowledgments} 

\noindent{This work was supported by Ministry of Science and
  Technology (Grant MOST 103-2112-M-001-016 \& 100-2811-M-001-109),
  NCTS (Taipei), ANCEF (Grant 2747), and the Pays de la Loire (Grant
  2010-11967).}

\newpage

\end{document}